\documentclass[12pt]{article}
\usepackage{a4wide}
\usepackage{latexsym}
\usepackage{amsmath}
\usepackage{amsfonts}
\usepackage{amscd}
\usepackage{cite}
\usepackage{graphicx}
\usepackage{axodraw}
\usepackage{float}

\usepackage{pslatex}
\usepackage[latin1]{inputenc}
\usepackage[OT2,T1]{fontenc}

\newcommand{\bq}{\begin{eqnarray}}
\newcommand{\eq}{\end{eqnarray}}
\newcommand{\eps}{\varepsilon}

\DeclareSymbolFont{cyrletters}{OT2}{wncyr}{m}{n}
\DeclareMathSymbol{\Sha}{\mathalpha}{cyrletters}{"58}

\begin{document}

\thispagestyle{empty}

\begin{flushright}
  MITP/14-108
\end{flushright}

\vspace{1.5cm}

\begin{center}
  {\Large\bf Fermions and the scattering equations\\
  }
  \vspace{1cm}
  {\large Stefan Weinzierl\\
\vspace{2mm}
      {\small \em PRISMA Cluster of Excellence, Institut f{\"u}r Physik, }\\
      {\small \em Johannes Gutenberg-Universit{\"a}t Mainz,}\\
      {\small \em D - 55099 Mainz, Germany}\\
  } 
\end{center}

\vspace{2cm}

% abstract -------------------------------------------------------------------------
\begin{abstract}\noindent
  {
This paper investigates how tree-level amplitudes with massless quarks, gluons and/or massless scalars transforming
under a single copy of the gauge group can be expressed 
in the context of the scattering equations as a sum over the inequivalent solutions of the scattering equations.
In the case where the amplitudes satisfy cyclic invariance, KK- and BCJ-relations
the only modification is the generalisation of the permutation invariant function $E(z,p,\eps)$.
We present a method to compute the modified $\hat{E}(z,p,\eps)$.
The most important examples are tree amplitudes in ${\mathcal N}=4$ SYM and
QCD amplitudes with one quark-antiquark pair and an arbitrary number of gluons.
QCD amplitudes with two or more quark-antiquark pairs do not satisfy the BCJ-relations
and require in addition a generalisation of the Parke-Taylor factors $C_\sigma(z)$.
The simplest case of the QCD tree-level four-point amplitude with two quark-antiquark pairs is discussed explicitly.
   }
\end{abstract}

\vspace*{\fill}

% main text ------------------------------------------------------------------------
\newpage

% ----------------------------------------------------------------------------------
\section{Introduction}
\label{sect:intro}

The scattering equations are a set of algebraic equations, which associate to the $n$ momentum vectors of a scattering event
$(n-3)!$ inequivalent $n$-tuples of complex numbers $z=(z_1,...,z_n)$.
These scattering equations have been studied in a series of papers by 
Cachazo, He and Yuang \cite{Cachazo:2013iaa,Cachazo:2013gna,Cachazo:2013hca,Cachazo:2013iea,Cachazo:2014nsa,Cachazo:2014xea}.
It is remarkable, that tree amplitudes for scalars (spin $0$), gluons (spin $1$) or gravitons (spin $2$)
can all be expressed elegantly as a sum over the $(n-3)!$ inequivalent solutions of the scattering equations.
Hereby, the scalar theory consists of scalars charged under two gauge groups $\mathrm{U}(N)$ and $\mathrm{U}(\tilde{N})$ with cubic interactions.
The full scalar amplitude is decomposed into ``double-partial'' amplitudes which are defined by a fixed colour order with
respect to $\mathrm{U}(N)$ as well as $\mathrm{U}(\tilde{N})$.
These double-partial amplitudes consist of Feynman diagrams compatible with both cyclic orders.
The statement above refers to the double-partial amplitudes.
In the gluon case it is well known that tree amplitudes can be decomposed 
into partial amplitudes with a fixed colour order (and hence with a fixed cyclic order).
The statement above refers in the gluonic case to the partial amplitudes.
Gravity amplitudes carry no additional gauge group information.
Therefore they are not cyclic ordered.
We see that while we increase the spin of the particles, the requirement on the cyclic ordering is relaxed.

The fact that tree amplitudes for scalars, gluons and gravitons 
can all be written as a sum over the solutions of the scattering equations has triggered significant interest in the community.
Here, we list only a few examples:
Dolan and Goddard \cite{Dolan:2013isa} have provided a proof that the amplitudes can be written in this form.
Their proof is based on 
Britto-Cachazo-Feng-Witten (BCFW) recursion relations \cite{Britto:2005fq}.
In addition, they showed how to rewrite the scattering equations in polynomial form \cite{Dolan:2014ega}.
The solutions are thus a (zero-dimensional) algebraic variety. This idea has been generalised in \cite{He:2014wua}.
There are close connections of the scattering equations 
with the construction of BCJ-numerators as shown in \cite{Monteiro:2013rya}.
Analytic solutions of the scattering equations have been investigated in \cite{Kalousios:2013eca,Weinzierl:2014vwa,Lam:2014tga}
and extensions to the scattering of massive particles are discussed in \cite{Dolan:2013isa,Naculich:2014naa}.
In addition, there are interesting connections with string theory \cite{Mason:2013sva,Berkovits:2013xba,Gomez:2013wza,Adamo:2013tsa,Geyer:2014fka,Casali:2014hfa}
and -- of course -- gravity \cite{Schwab:2014xua,Afkhami-Jeddi:2014fia,Zlotnikov:2014sva,Kalousios:2014uva,White:2014qia,Monteiro:2014cda}.

As mentioned above, the formalism based on the scattering equations describes nicely the scattering of particles
of spin $0$, $1$, and $2$.
What is missing in this list are particles of half-integer spin, in particular fermions of spin $1/2$.
Not much is known in this direction for non-supersymmetric theories.
In supersymmetric theories amplitudes can be described in terms of superfields, and amplitudes with external fermions are just specific components of
superamplitudes \cite{Roiban:2004yf,Cachazo:2013iaa}.
Furthermore, the four-fermion and the two-fermion-two-gluon amplitude were discussed explicitly 
in a supersymmetric string theory context \cite{Bjerrum-Bohr:2014qwa}.

In the pure gluonic case the scattering equations allow us to write the tree partial gluon amplitude as
\bq
\label{example_gluons}
 A_n\left(p, \eps, \sigma\right)
 & = &
 i
 \sum\limits_{\mathrm{solutions} \; j} J\left(z^{(j)},p\right) \; C_\sigma\left(z^{(j)}\right) \; E\left(z^{(j)},p,\eps\right).
\eq
The exact definition of all quantities appearing in this equation will be given in section~(\ref{scalars_gluons_gravitons}).
The quantity $J$ is a Jacobian factor, depending on the variables of the scattering equations $z$ and the external momenta $p$.
The Parke-Taylor factor $C_\sigma$ is a quantity depending only on $z$ and the cyclic order $\sigma$.
The quantity $E$ depends on $z$, $p$ and the external polarisations $\eps$, but does not depend on the cyclic order $\sigma$.
Eq.~(\ref{example_gluons}) can be interpreted as a ``factorisation of information'': 
The information on the cyclic ordering is contained in $C_\sigma$, the information on the external polarisation is contained in $E$.

QCD amplitudes involving quarks can be decomposed into primitive amplitudes with a fixed cyclic ordering.
The same can be done for amplitudes involving gluinos in ${\mathcal N}=4$ SYM.
In this paper we investigate cyclic ordered amplitudes with particles of spin up to $1$ and study under which conditions
a representation in the form as in eq.~(\ref{example_gluons}) with a modified function $\hat{E}$ exists.
Our first assumption is that neither the scattering equations (and hence the Jacobian) nor the Parke-Taylor factor
$C_\sigma$ are changed.
We show that a representation in the form of eq.~(\ref{example_gluons}) with a modified $\hat{E}$ exists 
if and only if the amplitudes satisfy cyclic invariance, the Kleiss-Kuijf (KK) relations \cite{Kleiss:1988ne}
and the Bern-Carrasco-Johansson (BCJ) relations \cite{Bern:2008qj}.
The amplitudes may contain as external particles a combination of spin $1$, spin $1/2$ particles and spin $0$ particles,
all charged under a single gauge group.
A typical example involving scalars are scalar particles in ${\mathcal N}=4$ SYM.
Note that the scalars in ${\mathcal N}=4$ SYM are not identical to the double-charged scalars
mentioned earlier.
In the case where a representation in the form of eq.~(\ref{example_gluons}) exists, we give a method to compute $\hat{E}$.
Our method expresses $\hat{E}$ as a linear combination of $(n-3)!$ basis amplitudes $A_n(\sigma)$.
Tree amplitudes in any theory defined by a Lagrangian can be computed easily by a variety of methods (Feynman diagrams,
Berends-Giele recursion relations \cite{Berends:1987me}, 
BCFW-recursion relations \cite{Britto:2005fq}).
Therefore we can consider tree amplitudes to be known quantities.
Our aim is not to find a computational method for an otherwise unknown quantity, in which case the approach of expressing
$\hat{E}$ in terms of the quantities we would like to calculate would be tautological.
The aim of this paper is to study whether a known quantity $A_n$ has a representation in the form of eq.~(\ref{example_gluons})
and to determine the function $\hat{E}$. The knowledge of the the function $\hat{E}$ is useful for generalised double-copy
relations.

The conditions for such a representation to exist are that the amplitudes fulfil cyclic invariance, the KK-relations and the BCJ-relations.
Tree amplitudes in ${\mathcal N}=4$ SYM satisfy these conditions and so do QCD amplitudes with a single quark-antiquark pair and 
an arbitrary number of gluons.
However, although primitive QCD amplitudes with two or more quark-antiquark pairs satisfy cyclic invariance and the KK-relations,
they do in general not satisfy the BCJ-relations.
Hence, they do not have a representation in the 
form of eq.~(\ref{example_gluons})\footnote{In the first version on the arxive it was erroneously claimed that they do satisfy the BCJ-relations.}.
The simplest example which does not have a representation in the form of eq.~(\ref{example_gluons}) is the QCD amplitude
$A_4(\bar{q},q,\bar{q}',q')$. 
The Parke-Taylor factor $C_\sigma$ encodes the set 
of relations consisting of cyclic invariance, KK-relations and BCJ-relations.
Amplitudes not satisfying the BCJ-relations require apart from a modification of $E$
a generalisation of the Parke-Taylor factors $C_\sigma$.
With the help of the pseudo-inverse of a matrix we derive a sufficient and necessary condition
for a modified Parke-Taylor factor $\hat{C}_\sigma$.
The simplest case of the QCD tree-level four-point amplitude $A_4(\bar{q},q,\bar{q}',q')$
with two quark-antiquark pairs is worked out explicitly.

This paper is organised as follows:
In section~\ref{sect_review} we review basic facts about the colour decomposition of tree-level amplitudes,
the scattering equations, relations satisfied by the amplitudes and the Kawai-Lewellen-Tye (KLT) orthogonality relations.
In section~\ref{sect_main} we present the generalisation for amplitudes satisfying cyclic invariance, KK- and BCJ-relations.
In this case only a modification of the function $E(z,p,\eps)$ is required.
Section~\ref{sect_main2} is devoted to amplitudes not satisfying the BCJ-relations. These require in addition a
modification of the Parke-Taylor factor $C_\sigma(z)$.
Finally, section~\ref{sect_conclusions} contains the conclusions.
For the convenience of the reader we included the cyclic ordered Feynman rules relevant to primitive amplitudes in a short appendix.

% ----------------------------------------------------------------------------------
\section{Review of basic facts}
\label{sect_review}

\subsection{Colour decomposition}

Amplitudes corresponding to theories with gauge groups may be decomposed into
group-theoreti\-cal factors multiplied by kinematic functions called partial amplitudes.
These partial amplitudes do not contain any colour information and are gauge invariant objects. 
Let us start the discussion with the case, where all particles belong to the adjoint representation of the 
gauge group.
This is the case for amplitudes in ${\mathcal N}=4$ SYM and for gluon amplitudes in QCD.
There are several possible choices for a basis in colour space. 
A possible choice is the colour-flow basis.
This choice is obtained by attaching a factor
\bq
 \sqrt{2} T^a_{ij}
\eq
to each external gluon and by using subsequently the Fierz identity 
to eliminate the adjoint indices.
With the definition of the 
``colour closed strings''
\bq
\label{colour_closed_string}
 c_{\mathrm{closed}}(g_1,...,g_n) 
 & = &
 \delta_{i_{1} j_{2}} \delta_{i_{2} j_{3}} 
 ... \delta_{i_{n} j_{1}}
\eq
we have the colour decomposition of the gluon tree amplitudes in QCD
as
\bq
\label{colour_decomp_pure_gluon}
 {\cal A}_{n}^{\mathrm{QCD}}(g_1,g_2,...,g_n) 
 & = & 
 \left(\frac{g}{\sqrt{2}}\right)^{n-2} 
 \sum\limits_{\sigma \in S_{n}/Z_{n}} 
 c_{\mathrm{closed}}\left(g_{\sigma_1},...,g_{\sigma_n}\right)
 A_{n}^{\mathrm{partial}}\left( g_{\sigma_1}, ..., g_{\sigma_n} \right),
\eq
where the sum is over all non-cyclic permutations of the external gluon legs.
The partial amplitudes $A_{n}^{\mathrm{partial}}$ are gauge-invariant and have a fixed cyclic order.
We note that theories with $\mathrm{SU}(N)$ gauge groups and $\mathrm{U}(N)$ gauge groups share the same partial amplitudes, since
$\mathrm{U}(1)$-gluons cannot couple to particles in the adjoint representation.
A similar decomposition as in eq.~(\ref{colour_decomp_pure_gluon}) exists for amplitudes ${\mathcal A}_{n}^{\mathrm{SYM}}$ in ${\mathcal N}=4$ SYM.

Let us now turn to QCD tree amplitudes with quarks.
We consider tree-level amplitudes with $n_q$ quarks, $n_q$ anti-quarks
and $n_g$ gluons in an $\mathrm{SU}(N)$-gauge theory.
The total number of external partons is therefore
\bq
 n & = & 2 n_q + n_g.
\eq
All partons are assumed to be massless.
We will call these amplitudes tree-level QCD amplitudes.
Here, the decomposition into gauge-invariant cyclic ordered objects, which are called primitive tree amplitudes, is more involved 
and consists of three step:
(i) elimination of identical flavours,
(ii) stripping of colour factors,
(iii) elimination of $\mathrm{U}(1)$-gluons.

(i) Elimination of identical flavours:
Amplitudes with quark-antiquark pairs of identical flavour can always be related to amplitudes,
where all quark-antiquark pairs have different flavours.
This is achieved by summing over all quark permutations.
An amplitude with $n_q$ quark-antiquark pairs can be written as
\bq
\label{identical_quarks}
\lefteqn{
{\mathcal A}_n^{\mathrm{QCD}}\left( \bar{q}_1, q_1, ..., \bar{q}_2, q_2, ..., \bar{q}_{n_q}, q_{n_q} \right)
 = } & & \nonumber \\
 & & 
 \sum\limits_{\sigma \in S(n_q)} \left( -1 \right)^{\sigma} 
 \left( \prod\limits_{j=1}^{n_q} \delta^{\mathrm{flav}}_{\bar{q}_j q_{\sigma(j)} } \right)
     {\mathcal A}_n^{\mathrm{non-id}}\left( \bar{q}_1, q_{\sigma(1)}, ..., \bar{q}_2, q_{\sigma(2)}, ..., \bar{q}_{n_q}, q_{\sigma(n_q)} \right).
\eq
Here, $(-1)^\sigma$ equals $-1$ whenever the permutation is odd and equals $+1$ if the permutation is even.
In ${\mathcal A}_n^{\mathrm{non-id}}$ each external quark-antiquark pair $(\bar{q}_j, q_{\sigma(j)})$ is connected by a continuous fermion line.
The flavour factor $\delta^{\mathrm{flav}}_{\bar{q}_j q_{\sigma(j)} }$ ensures that this combination is only taken into account, if
$\bar{q}_j$ and $ q_{\sigma(j)}$ have the same flavour.
In ${\mathcal A}_n^{\mathrm{non-id}}$ each external quark-antiquark pair $(\bar{q}_j, q_{\sigma(j)})$ is treated as having a flavour different from all other
quark-antiquark pairs.
It is therefore sufficient to discuss only the case of different quark flavours 
and we will therefore from now on assume that all quark flavours are different.

(ii) Stripping of colour factors:
As in the pure gluonic case we may decompose the amplitudes into group-theoretical factors 
(carrying the colour structures) multiplied by partial amplitudes.
As before we work in the colour-flow basis and attach a factor
$\sqrt{2} T^a_{ij}$
to each external gluon and use subsequently the $\mathrm{SU}(N)$-Fierz identity 
\bq
\label{fierz_identity}
 T^a_{ij} T^a_{kl} & = &  \frac{1}{2} \left( \delta_{il} \delta_{jk}
                         - \frac{1}{N} \delta_{ij} \delta_{kl} \right).
\eq
to eliminate the adjoint indices.
In addition to the ``colour closed string'' introduced in eq.~(\ref{colour_closed_string})
we also have to consider a
``colour open string''
\bq
 c_{\mathrm{open}}(q,g_1,...,g_n,\bar{q})
 & = &
 \delta_{i_q j_{1}} \delta_{i_{1} j_{2}} 
 ... \delta_{i_{n} j_{\bar{q}}}.
\eq
Born amplitudes with quarks have a decomposition in colour factors, which are products
of colour open strings. The colour decomposition of a Born amplitude with $n_q$ quarks, $n_q$ antiquarks and $n_g$ gluons 
reads \cite{Mangano:1990by}
\bq
\label{tree_multi_quark}
 {\cal A}_n^{\mathrm{non-id}} & = &
 \left(\frac{g}{\sqrt{2}}\right)^{n-2}
 \sum\limits_{\sigma \in S_{n_g}} \sum\limits_{\pi \in S_{n_q}}
 \sum\limits_{\stackrel{i_1,...,i_{n_q} \ge 0}{i_1+...+i_{n_q}=n_g}}
 c_{\mathrm{open}}\left(q_1,g_{\sigma_1},...,g_{\sigma_{i_1}},\bar{q}_{\pi_1}\right)
 \nonumber \\
 & & 
 c_{\mathrm{open}}\left(q_2,g_{\sigma_{i_1+1}},...,g_{\sigma_{i_1+i_2}},\bar{q}_{\pi_2}\right)
 ...
 c_{\mathrm{open}}\left(q_{n_q},g_{\sigma_{i_1+...+i_{n_q-1}+1}},...,g_{\sigma_{i_1+...+i_{n_q}}},\bar{q}_{\pi_{n_q}}\right)
 \nonumber \\
 & &
 A^{\mathrm{partial}}_n\left(q_1,g_{\sigma_1},...,g_{\sigma_{i_1}},\bar{q}_{\pi_1}, q_2, ..., g_{\sigma_{i_1+...+i_{n_q}}}, \bar{q}_{\pi_{n_q}} \right).
\eq
The sum over $\sigma$ is over all permutations of the external gluons,
the sum over $\pi$ is over all permutations of the colour indices of the antiquarks.
The sum over $\{i_1,...,i_{n_q}\}$ is over all partitions of $n_g$ into $n_q$ non-negative integers and corresponds to the different possibilities to
distribute $n_g$ gluons among $n_q$ open strings. 
The partial amplitudes $A^{\mathrm{partial}}_n$ are gauge-invariant, but are in general 
for $n_q > 2$ not cyclic ordered.
This is related to the fact that for $n_q \ge 2$ there can be so-called $\mathrm{U}(1)$-gluons, corresponding to the second term of the Fierz identity in eq.~(\ref{fierz_identity}).

(iii) Elimination of $\mathrm{U}(1)$-gluons:
The partial amplitudes may be further decomposed into smaller objects, called primitive amplitudes.
Primitive tree amplitudes are purely kinematic objects, which are gauge-invariant and which have a fixed cyclic ordering of the external
legs.
They are calculated from planar diagrams with the colour-ordered Feynman rules given in appendix \ref{appendix:colour_ordered_rules}.
These Feynman rules correspond to the colour-stripped Feynman rules of a $\mathrm{U}(N)$-gauge theory with quarks in the adjoint representation.
In a $\mathrm{U}(N)$-gauge theory additional $\mathrm{U}(1)$-gluons are absent, since the Fierz identity for $\mathrm{U}(N)$ simply reads
\bq
 T^a_{ij} T^a_{kl} & = &  \frac{1}{2} \delta_{il} \delta_{jk},
 \;\;\;\;\;\;
 \mbox{for $\mathrm{U}(N)$.}
\eq
With quarks in the adjoint representation, all colour-ordered three-valent vertices are anti-sym\-metric under the exchange of two of the three external particles.
In general, the linear combination, which expresses a partial amplitude in terms of primitive amplitudes is not unique.
This is due to the fact that there are in general relations among the primitive amplitudes.
The Kleiss-Kuijf relations are an example. 
There is a systematic combinatorial algorithm for the decomposition of partial multi-quark amplitudes into primitive amplitudes, given in \cite{Reuschle:2013qna}.
Any permutation may be written as a product of cycles. Let us assume that the permutation $\pi$ occurring in eq.~(\ref{tree_multi_quark})
consists of $r$ cycles. 
We denote the order of the $i$-th cycle by $k_i$.
Thus there are $k_i$ open strings associated to cycle $i$.
Each cycle $i$ defines a cyclic word $u_i$, obtained by the concatenation of the arguments of the $k_i$ open strings.
In this way we obtain $r$ cyclic words $u_1$, ..., $u_r$.
Ref.~\cite{Reuschle:2013qna} defines a shuffle operation $U(u_1,...,u_r)$
and the combinatorial decomposition of the partial amplitude into primitive amplitudes is given by
\bq
 A^{\mathrm{partial}}_n
 & = &
 \left(-\frac{1}{N}\right)^{r-1}
 \sum\limits_{w \in U\left(u_1, ..., u_r\right)} 
 A_n^{\mathrm{primitive}}\left( w \right),
\eq
where the sum is over all cyclic words appearing in the shuffle operation $U(u_1, ..., u_r)$.

For completeness we briefly discuss the colour decomposition of a cubic scalar theory with particles 
in the adjoint representation of $\mathrm{U}(N) \times \mathrm{U}(\tilde{N})$.
Here we have a double decomposition with respect to the groups $\mathrm{U}(N)$ and $\mathrm{U}(\tilde{N})$:
\bq
\label{double_colour_decomposition}
 \mathcal{S}_n(\phi_1,\phi_2,...,\phi_n)
 & = &
 \left(\frac{g}{2}\right)^{n-2} 
 \sum\limits_{\sigma, \tilde{\sigma} \in S_{n}/Z_{n}} 
 c_{\mathrm{closed}}\left(\phi_{\sigma_1},...,\phi_{\sigma_n}\right)
 c_{\mathrm{closed}}\left(\phi_{\tilde{\sigma}_1},...,\phi_{\tilde{\sigma}_n}\right)
 \nonumber \\
 & &
 S_{n}\left( \phi_{\sigma_1}, ..., \phi_{\sigma_n} | \phi_{\tilde{\sigma}_1}, ..., \phi_{\tilde{\sigma}_n} \right).
\eq
The partial amplitude $S_{n}(\phi_{\sigma_1}, ..., \phi_{\sigma_n} | \phi_{\tilde{\sigma}_1}, ..., \phi_{\tilde{\sigma}_n} )$
corresponds to Feynman diagrams compatible with the cyclic orders $\sigma$ and $\tilde{\sigma}$ \cite{Cachazo:2013iea}.

\subsection{The scattering equations}

We denote by $\Phi_n$ the momentum configuration space of $n$ external massless particles:
\bq
 \Phi_n & = &
 \left\{ \left(p_1,p_2,...,p_n\right) \in \left({\mathbb C} M\right)^n | p_1+p_2+...+p_n=0, p_1^2 = p_2^2 = ... = p_n^2 = 0 \right\}.
\eq
In other words, a $n$-tuple $p=(p_1, p_2, ..., p_n)$ of momentum vectors belongs to $\Phi_n$ if this $n$-tuple satisfies momentum conservation
and the mass-shell conditions $p_i^2=0$ for massless particles. It will be convenient to use the notation $p$  without any index
to denote such an $n$-tuple.

We further denote by $\hat{\mathbb C} = {\mathbb C} \cup \{\infty\}$.
The space $\hat{\mathbb C}$ is equivalent to the complex projective space ${\mathbb C}{\mathbb P}^1$.
For amplitudes with $n$ external particles we consider the space $\hat{\mathbb C}^n$.
Points in $\hat{\mathbb C}^n$ will be denoted by $z=(z_1,z_2,...,z_n)$.
Again we use the convention that $z$ without any index denotes an $n$-tuple.
We set for $1\le i \le n$
\bq
 f_i\left(z,p\right) & = & 
 \sum\limits_{j=1, j \neq i}^n \frac{ 2 p_i \cdot p_j}{z_i - z_j}.
\eq
Differences like in the denominator will occur quite often in this article and we use the abbreviation
\bq
 z_{ij} & = & z_i - z_j.
\eq
The scattering equations read \cite{Cachazo:2013hca}
\bq
\label{scattering_equations}
 f_i\left(z,p\right) & = & 0.
\eq
For a fixed $p \in \Phi_n$ a solution of the scattering equation is a point $z \in \hat{\mathbb C}^n$, such that the
scattering equations in eq.~(\ref{scattering_equations}) are satisfied.

The scattering equations are invariant under the projective special linear group
$\mathrm{PSL}(2,{\mathbb C})=\mathrm{SL}(2,{\mathbb C})/{\mathbb Z}_2$.
Here, ${\mathbb Z}_2$ is given by $\{ {\bf 1}, -{\bf 1} \}$, with ${\bf 1}$ denoting
the $(2 \times 2)$-unit matrix.
Let
\bq
 g = \left(\begin{array}{cc} a & b \\ c & d \\ \end{array} \right) & \in & \mathrm{PSL}(2,{\mathbb C}).
\eq
Each $g \in \mathrm{PSL}(2,{\mathbb C})$ acts on a single $z_i \in \hat{\mathbb C}$ as follows:
\bq
 g \cdot z_i & = &
 \frac{a z_i + b}{c z_i + d}.
\eq
We further set
\bq
 g \cdot \left(z_1, z_2, ..., z_n \right)
 & = &
 \left(g \cdot z_1, g \cdot z_2, ..., g \cdot z_n \right).
\eq
If $(z_1,z_2, ..., z_n)$ is a solution of eq.~(\ref{scattering_equations}), then also
$(z_1',z_2', ..., z_n') = g \cdot (z_1,z_2, ..., z_n)$ is a solution.
We call two solutions which are related by a $\mathrm{PSL}(2,{\mathbb C})$-transformation equivalent solutions.
We are in particular interested in the set of all inequivalent solutions of the scattering equations.
As shown in \cite{Cachazo:2013iaa,Cachazo:2013gna}, there are $(n-3)!$ different solutions not related by a $\mathrm{PSL}(2,{\mathbb C})$-transformation.
We will denote a solution by 
\bq
 z^{(j)} & = & \left( z_1^{(j)}, ..., z_n^{(j)} \right)
\eq
and a sum over the $(n-3)!$ inequivalent solutions by
\bq
 \sum\limits_{\mathrm{solution}\;j}
\eq
The $n$ scattering equations in eq.~(\ref{scattering_equations}) are not independent, only $(n-3)$ of them are.
The M\"obius invariance implies the relations
\bq
 \sum\limits_{j=1}^n f_j\left(z,p\right) = 0,
 \;\;\;\;\;\;
 \sum\limits_{j=1}^n z_j f_j\left(z,p\right) = 0,
 \;\;\;\;\;\;
 \sum\limits_{j=1}^n z_j^2 f_j\left(z,p\right) = 0.
\eq

\subsection{Amplitudes for scalars, gluons and gravitons}
\label{scalars_gluons_gravitons}

The scattering equations allow for an elegant representation of the scalar amplitude, the gluon amplitude
and the graviton amplitude \cite{Cachazo:2013hca,Cachazo:2013iea,Dolan:2013isa}.
In this paragraph we review these representations.
One first defines a $n \times n$-matrix $\Phi(z,p)$ with entries
\bq
 \Phi_{ab}\left(z,p\right)
 & = &
 \frac{\partial f_a\left(z,p\right)}{\partial z_b}
 \;\; = \;\;
 \left\{
 \begin{array}{cc}
 \frac{2 p_a \cdot p_b}{z_{ab}^2} & a \neq b, \\
 - \sum\limits_{j=1, j \neq a}^n \frac{2 p_a \cdot p_j}{z_{aj}^2} & a=b. \\
 \end{array}
 \right.
\eq
Let $\Phi^{ijk}_{rst}(z,p)$ denote the $(n-3)\times(n-3)$-matrix, where the rows $\{i,j,k\}$ and the columns $\{r,s,t\}$ have
been deleted.
We set
\bq
 \det{}' \; \Phi\left(z,p\right)
 & = &
 \left(-1\right)^{i+j+k+r+s+t}
 \frac{\left|\Phi^{ijk}_{rst}(z,p)\right|}{\left(z_{ij}z_{jk}z_{ki}\right)\left(z_{rs}z_{st}z_{tr}\right)}.
\eq
With the above sign included, the quantity
$\det{}' \; \Phi(z,p)$ is independent of the choice of $\{i,j,k\}$ and $\{r,s,t\}$.
One defines a Jacobian factor by
\bq
\label{def_jacobian}
 J\left(z,p\right) & = &
 \frac{1}{\det{}' \; \Phi\left(z,p\right)}.
\eq
One further defines a $(2n)\times(2n)$ antisymmetric matrix $\Psi(z,p,\eps)$, where 
$\eps = (\eps_1,...,\eps_n)$ denotes the $n$-tuple of external polarisation vectors, through 
\bq
\label{def_Psi_1}
 \Psi\left(z,p,\eps\right)
 & = &
 \left( \begin{array}{cc}
 A & - C^T \\
 C & B \\
 \end{array} \right)
\eq
with
\bq
 A_{ab}
 = 
 \left\{ \begin{array}{cc}
 \frac{2 p_a \cdot p_b}{z_{ab}} & a \neq b, \\
 0 & a = b, \\
 \end{array} \right.
 & &
 B_{ab}
 = 
 \left\{ \begin{array}{cc}
 \frac{2 \eps_a \cdot \eps_b}{z_{ab}} & a \neq b, \\
 0 & a = b, \\
 \end{array} \right.
\eq
and
\bq
\label{def_Psi_3}
 C_{ab}
 & = &
 \left\{ \begin{array}{cc}
 \frac{2 \eps_a \cdot p_b}{z_{ab}} & a \neq b, \\
 - \sum\limits_{j=1, j \neq a}^n \frac{2 \eps_a \cdot p_j}{z_{aj}}  & a = b. \\
 \end{array} \right.
\eq
For $1 \le i < j \le n$.
one denotes by $\Psi^{ij}_{ij}(z,p,\eps)$ the $(2n-2)\times(2n-2)$-matrix 
where the rows and columns $i$ and $j$ of $\Psi(z,p,\eps)$ have been deleted.
$\Psi^{ij}_{ij}(z,p,\eps)$ has a non-vanishing Pfaffian, which is independent of $i$ and $j$, and one sets
\bq
 E\left(z,p,\eps\right)
 & = &
 \frac{\left(-1\right)^{i+j}}{2 z_{ij}} \mathrm{Pf} \; \Psi^{ij}_{ij}\left(z,p,\eps\right).
\eq
Under a $\mathrm{PSL}(2,{\mathbb C})$ transformation we have
\bq
\label{PSL2_trafo_E}
 E\left(p,\eps,g \cdot z\right)
 & = &
 \left( \prod\limits_{j=1}^n \left(c z_j + d\right)^2 \right)
 E\left(p,\eps,z\right).
\eq
The function $E(z,p,\eps)$ is gauge-invariant and all $z$-variables transform with weight $2$ 
under $\mathrm{PSL}(2,{\mathbb C})$.
However, when we expand the Pfaffian the individual terms in this expansion are in general neither gauge-invariant nor
do they have the correct transformation properties with respect to $\mathrm{PSL}(2,{\mathbb C})$.
We will later see that there is an expansion into a sum of terms, such that each term in this sum is gauge-invariant and has the
correct $\mathrm{PSL}(2,{\mathbb C})$-transformation properties.

Let us now consider a permutation $\sigma \in S_n$.
One defines a cyclic factor (or Parke-Taylor factor) by
\bq
\label{def_cyclic_factor}
 C_\sigma\left(z\right)
 & = & 
 \frac{1}{z_{\sigma(1)\sigma(2)} z_{\sigma(2)\sigma(3)} ... z_{\sigma(n)\sigma(1)}}.
\eq
Under $\mathrm{PSL}(2,{\mathbb C})$ the Parke-Taylor factor transforms as
\bq
 C_\sigma\left(g \cdot z \right)
 & = &
 \left( \prod\limits_{j=1}^n \left(c z_j + d\right)^2 \right)
 C_\sigma\left(z\right).
\eq

We can now write the $n$-particle cyclic-ordered scalar amplitude $S_n$ with three-valent vertices, where we take for simplicity $\sigma=\tilde{\sigma}$ in eq.~(\ref{double_colour_decomposition}),
the $n$-gluon partial amplitude $A_n$ and
the $n$-graviton amplitude $M_n$
as \cite{Cachazo:2013hca,Cachazo:2013iea,Dolan:2013isa}
\bq
\label{scattering_amplitudes}
 S_n\left(p,\sigma\right)
 & = &
 i
 \sum\limits_{\mathrm{solutions} \; j} J\left(z^{(j)},p\right) \; \left[ C_\sigma\left(z^{(j)}\right) \right]^2,
 \nonumber \\
 A_n\left(p, \eps, \sigma\right)
 & = &
 i
 \sum\limits_{\mathrm{solutions} \; j} J\left(z^{(j)},p\right) \; C_\sigma\left(z^{(j)}\right) \; E\left(z^{(j)},p,\eps\right),
 \nonumber \\
 M_n\left(p, \eps \right)
 & = &
 i
 \sum\limits_{\mathrm{solutions} \; j} J\left(z^{(j)},p\right) \; \left[ E\left(z^{(j)},p,\eps\right) \right]^2.
\eq
We remark that the cyclic order $\sigma$ enters only through the cyclic factors $C_\sigma(z)$.
The Jacobian $J(z,p)$ and the function $E(z,p,\eps)$ are invariant under permutations.
The polarisations of the external states are entirely contained in the function $E(z,p,\eps)$.

Eq.~(\ref{scattering_amplitudes}) shows that the scattering equations can be used to express the scattering amplitudes
of spin 0, 1, and 2 particles.
This raises immediately the question, if also spin $1/2$ particles can be included in this representation.
We will investigate this question in this paper.

\subsection{Relations among amplitudes}

The pure gluonic tree amplitudes satisfy -- as the tree amplitudes in ${\mathcal N}=4$ SYM --
a set of equations, consisting of cyclic invariance relations, the Kleiss-Kuijf relations \cite{Kleiss:1988ne} 
and the Bern-Carrasco-Johansson relations \cite{Bern:2008qj}.
This set of equations allow us to express any amplitude out of the $n!$ possible external orderings 
as a linear combination of no more than $(n-3)!$ orderings.

Clearly, cyclic ordered amplitudes are cyclic invariant.
Cyclic invariance allows us to reduce the number of orderings from $n!$ to $(n-1)!$
and we may fix one external leg at a specific position, say external leg $1$ at position $1$.

The Kleiss-Kuijf relations allow us to fix a second external leg at a second specified position, say external
leg $2$ at position $2$.
This reduces the number of orderings to $(n-2)!$.
To state the Kleiss-Kuijf relations we let
\bq
 \vec{\alpha} = \left( \alpha_1, ..., \alpha_j \right),
 & & 
 \vec{\beta} = \left( \beta_1, ..., \beta_{n-2-j} \right)
\eq
and $\vec{\beta}^T = ( \beta_{n-2-j}, ..., \beta_1 )$.
The Kleiss-Kuijf relations read
\bq
\label{Kleiss_Kuijf}
 A_n\left( 1, \vec{\beta}, 2, \vec{\alpha} \right)
 & = & 
 \left( -1 \right)^{n-2-j}
 \sum\limits_{\sigma \in \vec{\alpha} \; \Sha \; \vec{\beta}^T}
 A_n\left( 1, 2, \sigma_1, ..., \sigma_{n-2} \right).
\eq
Here, $\vec{\alpha} \; \Sha \; \vec{\beta}^T$ denotes the set of all shuffles of $\vec{\alpha}$ with $\vec{\beta}^T$, i.e.
the set of all permutations of the elements of $\vec{\alpha}$ and $\vec{\beta}^T$, which preserve the relative order of the
elements of $\vec{\alpha}$ and of the elements of $\vec{\beta}^T$.

Finally, the BCJ relations allow to fix a third external leg at a third specified position, say external leg $3$ at position $3$.
This reduces the number of orderings to $(n-3)!$.
The BCJ relations have the form
\bq
\label{BCJ_relation}
 A_n\left( 1, 2, \vec{\beta}, 3, \vec{\alpha} \right)
 & = &
 \sum\limits_{\sigma \in POP\left(\vec{\alpha},\vec{\beta}\right)}
 {\mathcal F}\left(\sigma\right) \;
 A_n\left( 1, 2, 3, \sigma_1, ..., \sigma_{n-3} \right),
\eq
where the sum is over all permutations of the set $\vec{\alpha} \cup \vec{\beta}$, which preserve the relative order of $\alpha$.
The ${\mathcal F}(\sigma)$ are kinematical coefficients, given in \cite{Bern:2008qj}.

A sufficient, but not necessary condition for amplitudes to satisfy the KK relations and the BCJ relations
is the following criteria:
If for all pairs $(i,j)$ there exists a BCFW-shift such that the amplitudes fall off like $1/z$ if $i$ and $j$ are adjacent
and like $1/z^2$ if $i$ and $j$ are not adjacent, then the amplitudes satisfy the KK- and BCJ-relations.
The proof of this statement can be found in \cite{Feng:2010my}.
This proof is entirely formulated in quantum field theory and uses only the 
Britto-Cachazo-Feng-Witten (BCFW) recursion relations \cite{Britto:2005fq}.
It should be mentioned that historically the first proof of the BCJ relations 
has been derived from string theory \cite{BjerrumBohr:2009rd,Stieberger:2009hq}.

The pure gluonic tree amplitudes and amplitudes in ${\mathcal N}=4$ SYM have the required fall-off behaviour
under BCFW-shifts \cite{ArkaniHamed:2008yf,ArkaniHamed:2008gz,Schwinn:2007ee,Cheung:2008dn}
and therefore these amplitudes satisfy the KK- and BCJ-relations \cite{Feng:2010my,Jia:2010nz}.
For amplitudes in ${\mathcal N}=4$ SYM it is essential that not only the (bosonic) momentum components but also the 
(fermionic) Grassmann components \cite{ArkaniHamed:2008gz} are shifted.
In this case the BCFW recursion relation relates an amplitude to lower-point amplitudes with different external states.

The fact that the fall-off behaviour is sufficient but not necessary can be seen as follows:
Let us consider a primitive QCD amplitude $A_n^{\mathrm{primitive}}(\bar{q}_1,g_2,....,g_{n-1},q_n)$ with one quark-antiquark pair.
Under a $\bar{q}_1$-$q_n$-shift the amplitude goes at the best to a constant for $z \rightarrow \infty$.
However, the primitive amplitudes with one quark-antiquark pair and $(n-2)$-gluons satisfy the KK- and BCJ-relations.
This follows from the fact, that these primitive amplitudes are identical to amplitudes with one gluino-antigluino pair
and $(n-2)$ gluons in ${\mathcal N}=4$ SYM.

Now let us consider primitive QCD amplitudes $A_n^{\mathrm{primitive}}$ with possibly multiple quark-anti\-quark pairs.
These amplitudes satisfy the cyclic invariance relations and the KK-relations, but in general they do not satisfy the
BCJ-relations.
The proof that these amplitudes satisfy the KK-relations follows from the anti-symmetry of the vertices and can be found in
\cite{Reuschle:2013qna}.
In order to find a counter example of a primitive amplitude, which does not satisfy the BCJ-relations, we have
to look at primitive amplitudes with at least two quark-antiquarks pairs.
Primitive amplitudes with no quarks (i.e. the pure gluonic ones) as well as
the amplitudes with exactly one quark-antiquark pair satisfy the BCJ-relations, as we have seen.
The simplest counter example is given by $A_4^{\mathrm{primitive}}(q_1,\bar{q}_2,q_3',\bar{q}_4')$.
At four points the BCJ-relations read
\bq
\label{bcj_four_points}
 s_{23} A_4^{\mathrm{primitive}}(g_1,g_2,g_3,g_4)
 & = & 
 s_{24} A_4^{\mathrm{primitive}}(g_1,g_2,g_4,g_3),
 \nonumber \\
 s_{23} A_4^{\mathrm{primitive}}(q_1,\bar{q}_2,g_3,g_4)
 & = & 
 s_{24} A_4^{\mathrm{primitive}}(q_1,\bar{q}_2,g_4,g_3),
\eq
and hold for the $g_1,g_2,g_3,g_4$ and $q_1,\bar{q}_2,g_3,g_4$ amplitudes.
Eq.~(\ref{bcj_four_points}) does not hold for the $q_1$, $\bar{q}_2$, $q_3'$, $\bar{q}_4'$ QCD amplitudes.
Instead we have the (simpler) relation
\bq
 A_4^{\mathrm{primitive}}(q_1,\bar{q}_2,q_3',\bar{q}_4')
 & = & 
 - A_4^{\mathrm{primitive}}(q_1,\bar{q}_2,\bar{q}_4',q_3').
\eq
The KK-relations imply then
\bq
 A_4^{\mathrm{primitive}}(q_1,\bar{q}_4',\bar{q}_2,q_3') & = & 0,
\eq
e.g. a primitive amplitude with crossed fermion lines vanishes.

We remark that the ${\mathcal N}=4$ SYM amplitude with two distinct gluino-antigluino pairs satisfies the BCJ-relations
\bq
 s_{23} A_4^{\mathrm{SYM}}(q_1,\bar{q}_2,q_3',\bar{q}_4')
 & = & 
 s_{24} A_4^{\mathrm{SYM}}(q_1,\bar{q}_2,\bar{q}_4',q_3'),
\eq
as mentioned earlier. The amplitudes $A_4^{\mathrm{primitive}}(q_1,\bar{q}_2,q_3',\bar{q}_4')$ and
$A_4^{\mathrm{SYM}}(q_1,\bar{q}_2,q_3',\bar{q}_4')$ are not identical. The former is given by a single Feynman diagram consisting
of a gluon exchange in the $s$-channel, while the latter is given by two diagrams, the first as above and the second given by
a scalar exchange in the $t$-channel.

% ----------------------------------------------------------------------------------
\subsection{KLT orthogonality}

We will need one more technical ingredient.
The cyclic factors introduced in the previous paragraph satisfy an orthogonality relation
called Kawai-Lewellen-Tye (KLT) orthogonality \cite{Cachazo:2013gna,Kawai:1985xq,BjerrumBohr:2010ta,BjerrumBohr:2010hn}.

We introduce first some notation:
Let $\alpha \in S_n$ denote a permutation with 
\bq
 \alpha(1)=1, 
 \;\;\;\;\;\;
 \alpha(n-1)=n-1,
 \;\;\;\;\;\;
 \alpha(n)=n.
\eq
We denote the subgroup of all such permutations with a slight abuse of notation by $S_{n-3}$.
Similar, let $\beta \in S_n$ be a permutation with 
\bq
 \beta(1)=1, 
 \;\;\;\;\;\;
 \beta(n-1)=n, 
 \;\;\;\;\;\;
 \beta(n)=n-1.
\eq
The subgroup of all such permutations is denoted by $\bar{S}_{n-3}$.
Let $z^{(i)}=(z_1^{(i)},z_2^{(i)},...,z_n^{(i)})$ be a solution of the  scattering equations.
We write
\bq
 C_\alpha^i & = & C_\alpha\left(z^{(i)}\right),
 \;\;\;\;\;\; \alpha \in S_{n-3},
 \nonumber \\
 \bar{C}_\beta^i & = & C_\beta\left(z^{(i)}\right),
 \;\;\;\;\;\; \beta \in \bar{S}_{n-3}.
\eq
For fixed $i$ we denote by $C^i$ the vector of all $C^i_\alpha$ with $\alpha \in S_{n-3}$. 
The vector $C^i$ has the dimension $(n-3)!$, since there are $(n-3)!$ elements in $S_{n-3}$.
We have $(n-3)!$ such vectors, since there are $(n-3)!$ solutions of the scattering equations. 
We further denote by $\bar{C}^j$ the vector of all $C^j_\beta$ with $\beta \in \bar{S}_{n-3}$.
The vector $\bar{C}^j$ has as well the dimension $(n-3)!$, and there are again $(n-3)!$ such vectors.
An inner product is defined by
\bq
 \left( C^i, \bar{C}^j \right)
 & = &
 \sum\limits_{\alpha \in S_{n-3}} \sum\limits_{\beta \in \bar{S}_{n-3}}
  C^i_\alpha S\left[\alpha|\beta\right] \bar{C}^j_\beta,
\eq
with
\bq
 S\left[\alpha|\beta\right]
 & = &
 \left(-1\right)^n
 \prod\limits_{i=2}^{n-2} \left( 2 p_1 \cdot p_{\alpha(i)} + \sum\limits_{j=2}^{i-1} \theta_\beta\left(\alpha(j),\alpha(i)\right) \; 2 p_{\alpha(j)} \cdot p_{\alpha(i)} \right),
\eq
and 
\bq
\theta_\beta(\alpha(j),\alpha(i))
 & = & 
 \left\{
 \begin{array}{ll}
  1 & \mbox{if $\alpha(j)$  comes before $\alpha(i)$ in the sequence $(\beta(2),...,\beta(n-2))$}, \\
  0 & \mbox{otherwise}. \\
 \end{array}
 \right.
 \nonumber
\eq
One has \cite{Cachazo:2013gna}
\bq
\label{KLT_orthogonality_relation}
 \frac{\left( C^i, \bar{C}^j \right)}{\left( C^i, \bar{C}^i \right)^{\frac{1}{2}}\left( C^j, \bar{C}^j \right)^{\frac{1}{2}}}
 & = &
 \delta^{ij},
\eq
and
\bq
 \left( C^j, \bar{C}^j \right)
 & = &
 \frac{1}{J\left(z^{(j)},p\right)}.
\eq
Eq.~(\ref{KLT_orthogonality_relation}) is the KLT orthogonality relation.
Furthermore we have that the $(n-3) \times (n-3)$-matrix indexed by $\beta \in \bar{S}_{n-3}$ and
$\gamma \in S_{n-3}$
\bq
 \sum\limits_{\mathrm{solutions} \; j} \bar{C}^j_\beta J\left(z^{(j)},p\right) C^j_\gamma
\eq
is an inverse to $S[\alpha|\beta]$, as shown in \cite{Cachazo:2013iea}:
\bq
\label{inverse_KLT_matrix_relation}
 \sum\limits_{\beta \in \bar{S}_{n-3}}
 S\left[\alpha|\beta\right]
 \sum\limits_{\mathrm{solutions} \; j} \bar{C}^j_\beta J\left(z^{(j)},p\right) C^j_\gamma
 & = &
 \delta_{\alpha \gamma}.
\eq

% ----------------------------------------------------------------------------------
\section{Generalisation of the function $E$}
\label{sect_main}

Let us consider cyclic ordered amplitudes $A_n$, where the external particles have spin $\le 1$.
The external polarisations are now given by polarisation vectors $\eps_j$ for external gluons,
spinors $\bar{u}_j$ for out-going fermions, spinors $v_j$ for out-going anti-fermions and $1$ for external scalars.
For simplicity we will assume all particles to be out-going.
In order to keep the notation simple we will continue to denote the set of external polarisations
by $\eps$.
The cyclic order of the external particles will be denoted by $\sigma$.
For $n$ external particles there are $n!$ different cyclic orders.

We are interested in the question under which condition the amplitudes have a representation of the form
as in eq.~(\ref{example_gluons}):
\bq
\label{main_result}
 A_n\left(p, \eps, \sigma\right)
 & = &
 i
 \sum\limits_{\mathrm{solutions} \; j} J\left(z^{(j)},p\right) \; C_\sigma\left(z^{(j)}\right) \; \hat{E}\left(z^{(j)},p,\eps\right).
\eq
Here, the Jacobian $J$ and the cyclic factor $C_\sigma$ are identical to eq.~(\ref{def_jacobian}) and eq.~(\ref{def_cyclic_factor}), respectively.
We only allow at this stage a modification of the function $E$.
If such a representation exists, we would like to know $\hat{E}(z^{(j)},p,\eps)$.

We have the following theorem:
For a given set of external particles with spin $\le 1$ a representation in the form of eq.~(\ref{main_result})
exists if and only if the amplitudes with this set of external particles satisfy 
cyclic invariance, the KK-relations and the BCJ-relations.
In this case the modified function $\hat{E}(z^{(j)},p,\eps)$ is given by
\bq
\label{def_E_hat}
 \hat{E}\left(z^{(j)},p,\eps\right)
 & = &
 - i
 \sum\limits_{\alpha \in S_{n-3}} \sum\limits_{\beta \in \bar{S}_{n-3}}
 A_n\left(p, \eps, \alpha\right)
 \; S\left[\alpha|\beta\right] \; \bar{C}^j_\beta.
\eq
Note that each term in eq.~(\ref{def_E_hat}) is gauge invariant and has the correct transformation properties under
$\mathrm{PSL}(2,{\mathbb C})$.
As mentioned in the introduction we consider tree amplitudes
in any theory defined by a Lagrangian to be known quantities, which can be computed
by a variety of methods (Feynman diagrams, Berends-Giele recursion relations \cite{Berends:1987me}, 
BCFW-recursion relations \cite{Britto:2005fq}, etc.).
Therefore $\hat{E}(z^{(j)},p,\eps)$ can be computed as well.
Let us emphasize that the goal is not to find a new stand-alone method to compute $A_n$, but rather to investigate
under which condition a representation in the form of eq.~(\ref{main_result}) exists, and in the case it does, to find a method
to compute $\hat{E}(z^{(j)},p,\eps)$.

In order to prove the theorem let us first assume that the amplitudes $A_n$ satisfy cyclic invariance, KK-relations and BCJ-relations.
Let us define quantities $\hat{A}_n$ by
\bq
 \hat{A}_n\left(p, \eps, \sigma\right)
 & = &
 i
 \sum\limits_{\mathrm{solutions} \; j} J\left(z^{(j)},p\right) \; C_\sigma\left(z^{(j)}\right) \; \hat{E}\left(z^{(j)},p,\eps\right),
\eq
with $\hat{E}(z^{(j)},p,\eps)$ given by eq.~(\ref{def_E_hat}).
We have to show that
\bq
\label{claim}
 \hat{A}_n\left(p, \eps, \sigma\right)
 & = &
 A_n\left(p, \eps, \sigma\right)
\eq
for all $\sigma \in S_n$.

Eq.~(\ref{claim}) is easily proven for $\sigma \in S_{n-3}$:
We have with eq.~(\ref{inverse_KLT_matrix_relation})
\bq
\label{proof_S_n_3}
 \hat{A}_n\left(p, \eps, \sigma\right)
 & = &
 \sum\limits_{\mathrm{solutions} \; j} 
 \;
 \sum\limits_{\alpha \in S_{n-3}} 
 \; \sum\limits_{\beta \in \bar{S}_{n-3}}
 A_n\left(p, \eps, \alpha\right)
 \; S\left[\alpha|\beta\right] \; \bar{C}^j_\beta
 \; J\left(z^{(j)},p\right) \; C_\sigma^j
 \nonumber \\
 & = &
 \sum\limits_{\alpha \in S_{n-3}} 
 A_n\left(p, \eps, \alpha\right)
 \delta_{\alpha \sigma}
 \nonumber \\
 & = &
 A_n\left(p, \eps, \sigma\right).
\eq
In order to prove eq.~(\ref{claim}) for $\sigma \in S_n \backslash S_{n-3}$
it is the easiest to show that eq.~(\ref{main_result}) satisfies
cyclic invariance, the Kleiss-Kuijf-relations
and the Bern-Carrasco-Johansson relations.
These relations allow to express any of $n!$ possible external orderings in a basis
of $(n-3)!$ orderings. As such a basis we can take the $(n-3)!$ orderings, for which we have 
already proven our claim in eq.~(\ref{proof_S_n_3}).

The cyclic invariance is clear from an inspection of eq.~(\ref{main_result}).
The cyclic order $\sigma$ enters only through $C_\sigma^j$, which is cyclic invariant.

The Kleiss-Kuijf relations read (we drop the arguments $p$ and $\eps$ for the amplitudes)
\bq
 \hat{A}_n\left( 1, \vec{\beta}, 2, \vec{\alpha} \right)
 & = & 
 \left( -1 \right)^{n-2-j}
 \sum\limits_{\sigma \in \vec{\alpha} \; \Sha \; \vec{\beta}^T}
 \hat{A}_n\left( 1, 2, \sigma_1, ..., \sigma_{n-2} \right).
\eq
In our case this reduces to an algebraic relation between the cyclic factors $C_\sigma^j$
\bq
\label{KK_C_sigma}
 C^j_{\left( 1, \vec{\beta}, 2, \vec{\alpha} \right)}
 & = & 
 \left( -1 \right)^{n-2-j}
 \sum\limits_{\sigma \in \vec{\alpha} \; \Sha \; \vec{\beta}^T}
 C^j_{\left( 1, 2, \sigma_1, ..., \sigma_{n-2} \right)},
\eq
which is again easily verified.
We remark that eq.~(\ref{KK_C_sigma}) holds independently of $z^{(j)}$ being a solution of the scattering equations or not.

The BCJ relations reduce as well to a relation for the cyclic factors $C_\sigma^j$:
\bq
\label{BCJ_C_sigma}
 C^j_{\left( 1, 2, \vec{\beta}, 3, \vec{\alpha} \right)}
 & = &
 \sum\limits_{\sigma \in POP\left(\vec{\alpha},\vec{\beta}\right)}
 {\mathcal F}\left(\sigma\right) \;
 C^j_{\left( 1, 2, 3, \sigma_1, ..., \sigma_{n-3} \right)},
\eq
We now need that $z^{(j)}$ is a solution of the scattering equations.
However, eq.~(\ref{BCJ_C_sigma}) is independent of the flavour of the external particles and its validity can be inferred from the
pure gluonic case.
Therefore we have shown $\hat{A}_n(p,\eps,\sigma)=A_n(p,\eps,\sigma)$ for all $\sigma \in S_n$.

Now let us assume that the amplitudes $A_n$ do not satisfy the set of relations 
given by cyclic invariance, KK-relations and BCJ-relations. This means that there is at least one relation from this set, 
which is violated.
Let us write this relation as
\bq 
\label{violation}
 \sum\limits_{\sigma \in S_n} R\left(\sigma\right) A_n\left(p,\eps,\sigma\right) & \neq & 0.
\eq
Now let us further assume that $A_n$ has a representation in the form of eq.~(\ref{main_result}).
In this representation the dependence on the cyclic order enters only through the Parke-Taylor factors $C_\sigma^j$.
Any quantity, which has a representation of this form satisfies cyclic invariance, KK-relations and BCJ-relations.
Therefore
\bq 
 \sum\limits_{\sigma \in S_n} R\left(\sigma\right) A_n\left(p,\eps,\sigma\right) & = & 0,
\eq
which is a contradiction to eq.~(\ref{violation}).
Therefore a representation of this form cannot exist.
This completes the proof of the theorem.

Important examples, where a representation in the form of eq.~(\ref{main_result}) exists,
are -- apart from the pure gluonic tree amplitudes -- all tree amplitudes in 
${\mathcal N}=4$ SYM and the QCD tree amplitudes with one quark-antiquark pair and an arbitrary number of gluons.
In all these cases, the function $\hat{E}(z^{(j)},p,\eps)$ is given by eq.~(\ref{def_E_hat}).

Let us briefly consider MHV amplitudes with one quark-antiquark pair
\bq
 A_n(q_1^+,g_2^+,...,g_l^-,....,g_{n-1}^+,\bar{q}_n^-).
\eq
There is a SUSY Ward identity, which relates this amplitude to the corresponding all-gluon MHV amplitude:
\bq
 A_n(q_1^+,g_2^+,...,g_l^-,....,g_{n-1}^+,\bar{q}_n^-)
 & = &
 - \frac{\langle l 1 \rangle}{\langle l n \rangle}
 A_n(g_1^+,g_2^+,...,g_l^-,....,g_{n-1}^+,g_n^-).
\eq
It follows that the function $\hat{E}(z^{(j)},p,\eps)$ is given for the $q_1^+,g_2^+,...,g_l^-,....,g_{n-1}^+,\bar{q}_n^-$-case
on the scattering variety by
\bq
 \hat{E}(z^{(j)},p,\eps)
 & = &
 - \frac{\langle l 1 \rangle}{\langle l n \rangle}
 \;
 \frac{\left(-1\right)^{a+b}}{2 z_{ab}^{(j)}} \mathrm{Pf} \; \Psi^{ab}_{ab}\left(z^{(j)},p,\eps\right),
\eq
with $\Psi$ defined by eqs.~(\ref{def_Psi_1})-(\ref{def_Psi_3}).
Note that in $\Psi$ all external polarisations are gluon polarisation vectors.

% ----------------------------------------------------------------------------------
\section{Generalisation of the Parke-Taylor factor}
\label{sect_main2}

In the previous section we gave a sufficient and necessary condition for a set of amplitudes to have a representation
of the form~(\ref{main_result}).
This applies to the pure gluonic amplitudes, all amplitudes in 
${\mathcal N}=4$ SYM and the QCD tree amplitudes with one quark-antiquark pair and an arbitrary number of gluons.
The most notable exceptions are the QCD primitive tree amplitudes with two or more quark-antiquark pairs.
These amplitudes do in general not satisfy the BCJ relations.
The simplest example of such an exception is the four-point amplitude
$A_4^{\mathrm{primitive}}(q_1,\bar{q}_2,q_3',\bar{q}_4')$.
In order to present these amplitudes as a sum over the solution of the scattering equation we will need a further
generalisation of eq.~(\ref{example_gluons}).
Since the Parke-Taylor factors $C_\sigma$ have the set of cyclic invariance, KK- and BCJ-relations
buried in them, we have to look at generalisations of the Parke-Taylor factors.
Let us denote the information on the flavours of the external particles collectively by $f$.
We consider a representation of the form
\bq
\label{main_result2}
 A_n\left(p, \eps, \sigma\right)
 & = &
 i
 \sum\limits_{\mathrm{solutions} \; j} J\left(z^{(j)},p\right) \; \hat{C}_\sigma\left(z^{(j)},f\right) \; \hat{E}\left(z^{(j)},p,\eps\right).
\eq
Note that this form keeps as much as possible the original structure and can again be interpreted as a ``factorisation of information'':
The information on the external polarisations enters only through $\hat{E}$, the information on the cyclic order only through $\hat{C}$.
The information on the flavours of the external particles enters $\hat{E}$ and $\hat{C}$.
The Jacobian $J$ is defined as before.
We will first investigate under which conditions such a representation may exist.
In order to simplify the discussion we introduce the following short-hand notation:
We define a $n!$-dimensional vector $A_\sigma$ with components
\bq
 A_\sigma & = & A_n\left(p, \eps, \sigma\right),
\eq
a $n! \times (n-3)!$-dimensional matrix $M_{\sigma j}$ by
\bq
 M_{\sigma j}
 & = &
 J\left(z^{(j)},p\right) \; \hat{C}_\sigma\left(z^{(j)},f\right),
\eq
and a $(n-3)!$-dimensional vector $\hat{E}_j$ by
\bq
 \hat{E}_j & = & \hat{E}\left(z^{(j)},p,\eps\right).
\eq
Then eq.~(\ref{main_result2}) may be written compactly as
\bq
 A_\sigma & = & i \; M_{\sigma j} \hat{E}_j,
\eq
where a sum over $j$ is understood.
Let us further denote a pseudo-inverse matrix of $M_{\sigma j}$ by $N_{j \sigma}$.
$N_{j \sigma}$ is a $(n-3)! \times n!$-dimensional matrix with the property
\bq
 M N M \;\; = \;\; M,
 & &
 N M N \;\; = \;\; N.
\eq
From the theory of pseudo-inverse matrices \cite{Ben-Israel,Koecher} we have the following statement:
A representation in the form of eq.~(\ref{main_result2}) exists if and only if
\bq
\label{condition_pseudo_inverse}
 M_{\sigma j} N_{j \tau} A_\tau & = & A_\sigma,
\eq
in other words $A_\sigma$ must be an eigenvector of $M N$ with eigenvalue $1$.
In this case $\hat{E}_j$ is given by
\bq
\label{pseudo_inverse_solution}
 \hat{E}_j 
 & = &
 - i N_{j \sigma} A_\sigma - i \left( \delta_{jk} - N_{j\tau} M_{\tau k} \right) u_k,
\eq
where $u_k$ is an arbitrary $(n-3)!$-dimensional vector.
In the case, where $M_{\sigma j}$ has rank $(n-3)!$, a pseudo-inverse is given by
\bq
 N & = & \left( M^T M \right)^{-1} M^T
\eq
and we further have $N_{j \tau} M_{\tau k} = \delta_{j k}$. Then eq.~(\ref{pseudo_inverse_solution})
reduces to
\bq
\label{pseudo_inverse_solution2}
 \hat{E}_j 
 & = &
 - i N_{j \sigma} A_\sigma.
\eq
Eq.~(\ref{condition_pseudo_inverse}) allows us to verify or falsify an ansatz for $\hat{C}_\sigma$.
With a valid ansatz for $\hat{C}_\sigma$, the function $\hat{E}$ is then given by
eq.~(\ref{pseudo_inverse_solution}) or eq.~(\ref{pseudo_inverse_solution2}).

Let us now look at the example
$A_4^{\mathrm{primitive}}(q_1,\bar{q}_2,q_3',\bar{q}_4')$.
We set
\bq
\hat{C}_\sigma\left(z,f\right)
 & = &
 \left\{
 \begin{array}{rl}
 \frac{1}{z_{12} z_{23} z_{34} z_{41}}, & \mbox{if the cyclic order is $q_1,\bar{q}_2,q_3',\bar{q}_4'$,} \\
 -\frac{1}{z_{12} z_{23} z_{34} z_{41}}, & \mbox{if the cyclic order is $q_1,\bar{q}_2,\bar{q}_4',q_3'$,} \\
 0, & \mbox{if the cyclic order is $q_1,q_3',\bar{q}_2,\bar{q}_4'$.} \\
 \end{array}
 \right.
\eq
With
\bq
 \hat{E}\left(z,p,\eps\right)
 & = &
 - i \; \frac{1}{z_{12} z_{24} z_{43} z_{31}}
 \; 
 s_{12}
 \;
 A_4^{\mathrm{primitive}}(q_1,\bar{q}_2,q_3',\bar{q}_4')
\eq
we then have
\bq
 A_4^{\mathrm{primitive}}\left(p, \eps, \sigma\right)
 & = &
 i
 \sum\limits_{\mathrm{solutions} \; j} J\left(z^{(j)},p\right) \; \hat{C}_\sigma\left(z^{(j)},f\right) \; \hat{E}\left(z^{(j)},p,\eps\right)
\eq
for all $\sigma \in S_4$.

% ----------------------------------------------------------------------------------
\section{Conclusions}
\label{sect_conclusions}

In this paper we discussed the extension
of the representation of a cyclic ordered tree amplitude in the framework of the scattering equations
from the pure gluonic case to the case, where the external particles are allowed to have spin $\le 1$.
In the case where the amplitudes satisfy cyclic invariance, the KK- and BCJ-relations
the only required modification is the generalisation of the
permutation invariant function $E(z,p,\eps)$.
The most important examples are tree amplitudes in ${\mathcal N}=4$ SYM and
QCD amplitudes with one quark-antiquark pair and an arbitrary number of gluons.
We have presented a method to compute the modified function $E(z,p,\eps)$.
The modified function is given as a linear combination of $(n-3)!$ basis amplitudes.
Each term in this linear combination is gauge-invariant and has the correct transformation properties under
$\mathrm{PSL}(2,{\mathbb C})$-transformations.
We then moved on towards amplitudes not satisfying the BCJ-relations.
This concerns QCD amplitudes with two or more quark-antiquark pairs.
These amplitudes require in addition a generalisation of the Parke-Taylor factors.
We derived a sufficient and necessary condition for such a generalisation 
and worked out the simplest case of the QCD tree-level four-point amplitude with two quark-antiquark pairs explicitly.
Finding the correct generalised Parke-Taylor factors for all QCD multi-quark primitive amplitudes is a direction for future research.

% ----------------------------------------------------------------------------------
\begin{appendix}

\section{Feynman rules}
\label{appendix:colour_ordered_rules}

In this appendix we give a list of the colour ordered Feynman rules. 
They are obtained from the standard Feynman rules by extracting from each formula the coupling
constant and the colour part.
The propagators for quark and gluon particles are given by
\bq
\begin{picture}(85,20)(0,5)
 \ArrowLine(70,10)(20,10)
\end{picture} 
 & = &
 i\frac{k\!\!\!/+m}{k^2-m^2},
 \nonumber \\
\begin{picture}(85,20)(0,5)
 \Gluon(20,10)(70,10){-5}{5}
\end{picture} 
& = &
 \frac{-ig^{\mu\nu}}{k^2}.
\eq
The colour ordered Feynman rules for the three-gluon and the four-gluon vertices are
\bq
\begin{picture}(100,35)(0,50)
\Vertex(50,50){2}
\Gluon(50,50)(50,80){3}{4}
\Gluon(50,50)(76,35){3}{4}
\Gluon(50,50)(24,35){3}{4}
\LongArrow(56,70)(56,80)
\LongArrow(67,47)(76,42)
\LongArrow(33,47)(24,42)
\Text(60,80)[lt]{$k_1^{\mu_1}$}
\Text(78,35)[lc]{$k_2^{\mu_2}$}
\Text(22,35)[rc]{$k_3^{\mu_3}$}
\end{picture}
 & = &
 i \left[ g^{\mu_1\mu_2} \left( k_1^{\mu_3} - k_2^{\mu_3} \right)
         +g^{\mu_2\mu_3} \left( k_2^{\mu_1} - k_3^{\mu_1} \right)
         +g^{\mu_3\mu_1} \left( k_3^{\mu_2} - k_1^{\mu_2} \right)
   \right],
 \nonumber \\
 \nonumber \\
 \nonumber \\
\begin{picture}(100,35)(0,50)
\Vertex(50,50){2}
\Gluon(50,50)(71,71){3}{4}
\Gluon(50,50)(71,29){3}{4}
\Gluon(50,50)(29,29){3}{4}
\Gluon(50,50)(29,71){3}{4}
\Text(72,72)[lb]{\small $\mu_1$}
\Text(72,28)[lt]{\small $\mu_2$}
\Text(28,28)[rt]{\small $\mu_3$}
\Text(28,72)[rb]{\small $\mu_4$}
\end{picture}
 & = &
  i \left[
        2 g^{\mu_1\mu_3} g^{\mu_2\mu_4} - g^{\mu_1\mu_2} g^{\mu_3\mu_4} 
                                        - g^{\mu_1\mu_4} g^{\mu_2\mu_3}
 \right].
 \nonumber \\
 \nonumber \\
\eq
The Feynman rule for the quark-gluon vertex is given by
\bq
\label{colour_ordered_quark_gluon_antiquark_vertex}
\begin{picture}(100,35)(0,50)
\Vertex(50,50){2}
\Gluon(50,50)(80,50){3}{4}
\ArrowLine(50,50)(29,71)
\ArrowLine(29,29)(50,50)
\Text(82,50)[lc]{$\mu$}
\end{picture}
 \;\; = \;\;
 i \gamma^{\mu},
 & \;\;\;\;\;\;\;\;\; &
\begin{picture}(100,35)(0,50)
\Vertex(50,50){2}
\Gluon(50,50)(20,50){3}{4}
\ArrowLine(50,50)(71,71)
\ArrowLine(71,29)(50,50)
\Text(18,50)[rc]{$\mu$}
\end{picture}
 \;\; = \;\;
 -i \gamma^{\mu}.
 \nonumber \\
 \nonumber \\
\eq

\end{appendix}

% ----------------------------------------------------------------------------------
% references
\bibliography{/home/stefanw/notes/biblio}
\bibliographystyle{/home/stefanw/latex-style/h-physrev5}

\end{document}